\documentclass[pra,amsmath,amssymb,superscriptaddress]{revtex4-1}
\usepackage{graphicx,float}
\usepackage{hyperref}
\usepackage{soul,ulem}
\usepackage{xcolor}
\usepackage[utf8]{inputenc}
\usepackage{amsmath,amssymb}
\usepackage{mathrsfs}

\usepackage{tabularx}
\usepackage{algorithm}
\usepackage[noend]{algpseudocode}

\newcommand{\dn}{{\delta\mkern-1.0mu n}} 
\newcommand{\drho}{{\delta\mkern-1.0mu\rho}} 
\newcommand{\dwrho}{{\delta\widetilde{\rho}}}
\newcommand{\dU}{{\delta\mkern-1.0mu U}}
\newcommand{\dV}{{\delta\mkern-1.0mu V}}
\newcommand{\dH}{{\delta\mkern-1.0mu H}}

\def\A{\mathscr{A}}
\def\B{\mathscr{B}}
\def\C{\mathscr{C}}

\def\V{\mathbb{V}}
\def\W{\mathbb{W}}

\renewcommand{\i}{\ensuremath{\text{\normalfont I}}}
\newcommand{\ii}{\ensuremath{\text{\normalfont I\!I}}}

\def\ve{\varepsilon}

\def\dt{\Delta t}
\def\wt{\widetilde}

\def\E{\mathcal{E}}

\begin{document}
\preprint{}

\title{A Projection-based Reduced-order Method for Electron Transport Problems with Long-range Interactions}

\author{Weiqi Chu}
\email{Chu@math.ucla.edu }
\affiliation{Department of Mathematics, University of California, Los Angeles, CA 90095, USA.}

\author{Xiantao Li}
\email{Xiantao.Li@psu.edu}
\affiliation{Department of Mathematics, the Pennsylvania State University, University Park, PA 16802, USA}%

\begin{abstract}

Long-range interactions play a central role in electron transport. At the same time, they present a challenge for direct computer simulations, since sufficiently large portions of the bath have to be included in the computation to accurately compute the Coulomb potential.  This article presents a reduced-order approach, by deriving an open quantum model for the reduced density-matrix.  To treat the transient dynamics, the problem is placed in a reduced-order framework. The dynamics, described by the Liouville von Neumann equation, is projected to subspaces using a Petrov-Galerkin projection.  In order to recover the global electron density profile  as a vehicle to compute the Coulomb potential,  we propose a domain decomposition approach,  where the computational domain also includes segments of the bath that are selected using logarithmic grids. This approach leads to a multi-component self-energy that enters the effective Hamiltonian.  We demonstrate the accuracy of the reduced model using a molecular junction built from a Lithium chains.      
\end{abstract}

\maketitle

\section{Introduction}
The availability of high-performance computing (HPC)
has prompted substantial progress in electron structure calculations using sophisticated models. Tremendous efforts have been made to implement the time-dependent density-functional theory (TDDFT) \cite{runge1984density,ullrich2011time}  in the context of electron transport \cite{stefanucci2004time,kurth2005time,stefanucci2004time-prb,cheng2006simulating,sanchez2006molecular,varga2011time,taylor2001ab}. Numerous software packages have been developed to facilitate such effort \cite{tancogne2020octopus,SCC-DFTB}. 
The main computational challenge in such an approach is still the computational cost. A typical situation is a molecular junction, connected to two leads that can be regarded as quantum bath \cite{reed1999molecular}. Ideally, one should consider a sufficiently large quantum bath to correctly model the effect of the semi-infinite leads. This necessarily introduces a large number of electronic degrees of freedom to the system. In terms of a TDDFT model, this induces many more wave functions and more importantly, their representations at many more grid points into the computation. Specifically, the degrees of freedom  has roughly an $\mathcal{O}(N^2)$ scaling where $N$ denotes the number of electrons. In a crude tight-binding approximation, a relatively larger number of electrons can be considered. But the computational challenge still remains.

Numerous attempts have been made to construct models so that only the electronic degrees of freedom near the junction are involved in the computation. These efforts range from the theoretical approach to derive open quantum system models \cite{burke2005density}, to absorbing boundary conditions (ABC) \cite{baer2004ab,xie2014complex,li2019absorbing},  the reduced density-matrix approach by Subotnik et al.  \cite{subotnik2009nonequilibrium} the driven Liouville von Neumann (DLvN) approach \cite{zelovich2014state,zelovich2015molecule,zelovich2016driven,zelovich2017parameter}, to the reduced-order approach \cite{chu2020reduced}.  In principle, all these methodologies lead to a reduced quantum model with much fewer variables.

This paper focuses on such a reduction scheme for systems with Coulomb interactions. Coulomb interactions play a central role in electron transport and are responsible for the important observation of Coulomb blockade \cite{kurth2010dynamical,beenakker1991theory}. However, as suggested in \cite{antoine2008review} in the context of ABCs, it is still an open challenge to derive a boundary condition that takes into account such long-range effects. The ABCs can only be applied with a leap of faith. The development of the driven LvN approach \cite{zelovich2014state,zelovich2015molecule} and the reduced-order method \cite{chu2020reduced} also neglected this effect.  However, in a more careful approximation of TDDFT, {\it e.g.}, the real-space approximation \cite{beck2000real,marques2003} or the self-consistent tight-binding approximation \cite{SCC-DFTB}, one needs to take into account the Coulomb interactions, which has to be computed at each time step and it leads to considerable computational overhead.

In order to compute the Coulomb potential at the junction, we replace a direct truncation of the bath regions with a sparse representation, where the grid spacings are gradually increased. The use of non-uniform grids that concentrate more at the regions of interest is not foreign in quantum mechanics: It has been a popular technique for computing the radial component of a Schr\"{o}dinger equation, {\it e.g.}, the logarithmic grids in the computation of pseudo-potentials \cite{willand2013norm}.  
Overall, our selected representations form a `multi-connected' sub-domain. Following the previously developed reduced-order technique \cite{chu2020reduced}, we project the Liouville von Neumann equation into appropriate subspaces. As a result, the effective Hamiltonian consists of a self-energy that couples the sub-domains. The reduced density-matrix (RDM), with a function interpolation, provides the global electron density profile, which can subsequently be used to estimate the Coulomb potential at the junction.  The treatment using the domain decomposition method is reminiscent of the partition density-functional theory \cite{elliott2010partition}. However, the current approach targets the dynamics at the center region, rather than the electron density in the entire domain. 

The fact that the wave functions from density-functional theory are often extended creates a challenge for the model reduction effort based on the wave functions. Although such ideas have been discussed  \cite{kurth2005time,ullrich2011time} to reduce the Kohn-Sham orbitals to a subdomain, it is not clear how to derive a reduced model that involves fewer orbitals. Therefore, we will base our reduction approach on the Liouville von Neumann equation for the density-matrix.
A convenient and robust approach to derive a reduced model, as suggested by reduced-order techniques \cite{antoulas2010interpolatory,freund2000krylov,bai2002krylov}, is by Petrov-Galerkin projection, which reduces the method to finding appropriate subspaces.  
The reduced model for the RDM will be referred to as the reduced Liouville von Neumann equation (RLvN). An important observation is that the Lie group structure is no longer present, due to the fact that the effective Hamiltonian is non-Hermitian. Rather, the RLvN should be viewed as an open quantum system \cite{breuer2002theory}, where the influence of the bath is implicitly incorporated. An important difference from those models, {\it e.g.}, the Lindblad equation \cite{lindblad1976generators}, is that the present model does not have the trace-preserving property due to the electron transport nature. 

In Section \ref{sec: I}, we will outline the domain decomposition and model reduction methods to derive the RLvN. Details regarding the implementations of the reduced model are presented in Sections \ref{sec: alg1}--\ref{sec: alg3}. Results from several numerical experiments are demonstrated in Section \ref{sec: II}. 

\section{Methods and Algorithms}\label{sec: I}

 
\subsection{The Liouville von Neumann equation and its perturbation}
We formulate our method and algorithms based on the LvN equation,
\begin{equation}\label{eq: lvn}
    i \partial_t \rho = [H, \rho].
\end{equation}
Here $\rho(x, x', t)$ represents the density matrix of the entire system. Following TDDFT models \cite{marques2006time,ullrich2011time}, the Hamiltonian operator $H$ will depend on the electron density $n(x,t)$, which can be determined from the density-matrix $\rho$. These properties are expressed as follows,
\begin{equation}
    H(t)= H[n(t)], \quad n(x, t)= \rho(x, x, t).
\end{equation}


Although our model works with the density-matrix, our primary interest is in the electric current induced by a time-dependent external potential. The potential bias is denoted here by $\dU(t)$ and it is switched on at $t=0_+$. It triggers  the time evolution of electron density $n(r,t)$, which has an implicit influence on the Hamiltonian $H$ of the entire system.
Motivated by the theory of linear response \cite{gross1985local,dobson1997time,gunnarsson1976exchange}, we first incorporate the external potential and write the total Hamiltonian as follows,  
 \begin{equation}
  H_\text{tot}(t)=H[n(t)]+\dU(t).
 \end{equation} 
The response of the system due to the external potential in the linear response regime can be represented in terms of the {\it perturbed} density matrix,
\begin{equation}\label{eq: drho0}
 	\drho(t) :=\rho(t) - \rho_{0}, \quad \drho(0)=0.
\end{equation}

Here $\rho_0$ is the ground state density-matrix of the system in the {\it absence} of the biased potential. A variety of methods are available for computing the ground state \cite{martin2004@book,elliott2010partition}.  The corresponding electron density is denoted here by  $n_0(r)$,   with its perturbation  $\dn(r,t)$ given by,
\begin{equation}
\dn = \textrm{diag}\;\big( \drho\big).
\end{equation}

With the ground state density, we define an {\it unperturbed} Hamiltonian as
\begin{equation}
    H_0= -\frac12{\nabla^2} + V_H[n_0] + V_{XC}[n_0] +  V_{ion},
\end{equation}
where  $V_{XC}$ is the exchange correlation potential. $V_H$ is the Hartree potential, and it is expressed as follows,
\begin{equation}\label{eq: vH}
    V_H[n](x) = \int_{\mathbb{R}} v(x-x') n(x') dx' .
\end{equation}

As a result, the total Hamiltonian $H_\text{tot}(t)$ can be split up as follows,
\begin{equation}
    H_\text{tot}(t)= H_0 + \dH(t),
\end{equation}
where,
\begin{equation}\label{eq: dH0}
    \dH(t) = H[n_0 + \dn(t)] - H_0 + \dU(t).
\end{equation}

Traditionally, the perturbation of the Hamiltonian operator is treated under the linear-response framework \cite{casida1995time}, which is also known as the Sternheimer's formalism, which linearizes the Hamiltonian as follows,
\begin{equation}
 H[n_0 + \dn] - H_0  = \mathcal{K} \dn + \mathcal{O}(\dn^2).
 \end{equation}
Here the integral operator $\mathcal{K}$  is given in terms of the functional derivative with respective to the density,
\begin{equation}
    \mathcal{K} \dn := \int \frac{\delta V}{\dn(x')}[n_0(x)]  \dn(x') dx'.
\end{equation}

Although the Sternheimer's approach is important in determining electro-magnetic properties, it leads to a dense matrix due to the Coulomb term. Therefore,  we will keep the nonlinear form \eqref{eq: dH0} instead. As we will demonstrate,  it does not complicate the numerical implementation.  We will break $\dH$ in \eqref{eq: dH0} into three contributions, 
\begin{equation}\label{eq: dH}
    \dH = \dU + V_H[\dn] +  \left(V_{XC}[n_0+\dn] - V_{XC}[n_0]\right).
\end{equation}

Notice that the Hartree term is linear with respect to the charge density:
\begin{equation}
    V_H[\dn] =V_H[n] - V_H[n_0].
\end{equation}
Meanwhile, the exchange correlation potential is nonlinear. This nonlinearity will be kept in our formulation.

After substituting \eqref{eq: drho0} into \eqref{eq: lvn}, and dropping the high order term, $[\dH,\drho]$, the perturbed density matrix satisfies a response equation, given by,
\begin{equation}\label{eq: drhoeq}
 	i
 	\drho_t(t)= [H_0,\drho(t)] + \Theta(t).
\end{equation}
The perturbation approach also leads to a non-homogeneous term,
\begin{equation}
\Theta(t) = [\dH(t),\rho_{0}],
\end{equation}
which incorporates the influence of the external potential as well as the density fluctuation $\dn$.

\subsection{Reduced-order modeling}

To better elaborate the method, we first establish some notations. To begin with, we use $\Omega$ to denote the entire region. As a concrete example, we first discretize the Hamiltonian operator using a real space method  and the domain $\Omega$ consists of grid points. We denote the dimension by $N=\text{dim}(\Omega)$, and it indicates the dimension of the full system. In the case when a tight-binding method, {\it e.g.},\cite{ hourahine2020dftb+}, is used, the domain $\Omega$ consists of the atoms. 
The setup of the molecular junctions naturally divides the domain $\Omega$ into a {\it center region}, denoted here by $\Omega_\C$, and a {\it bath region}, denoted by $\Omega_\B$; $\Omega= \Omega_\B \cup \Omega_\C$ and $ \Omega_\B \cap \Omega_\C=\emptyset$. 

\begin{figure}[tph]
	\includegraphics[scale=0.52]{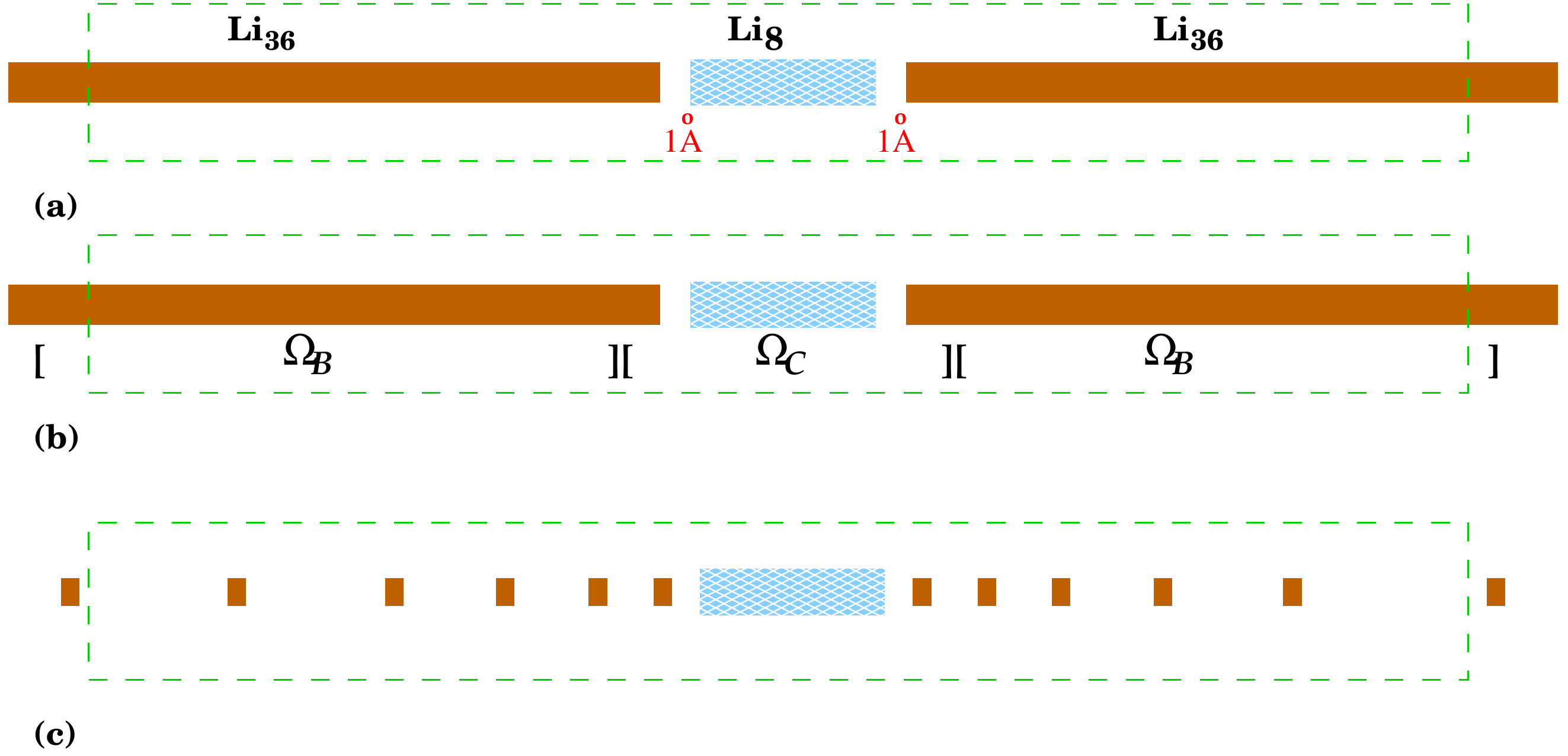}
	\caption{Schematic representation of the molecular junction and its representation in the computation. (a) The system with three Li chains separated by an extra spacing of 1\AA. (b) The domain decomposition in the bath and center regions. (c) An illustration of the selected grid points in the bath region with increasing spacings.}
	\label{fig: schematic2}
\end{figure}

As alluded to in the introduction, the main obstacle in the model reduction is that the Coulomb potential is non-local, and  completely neglecting the electron density in the bath region will lead to inaccurate computation of the Hartree potential.  To address this issue, we extend the `center' region $\Omega_\C$ by including points in the bath that are distributed according to,
\begin{equation}\label{eq: log-grid}
\begin{aligned}
       x_\alpha &= x_R +  a b^\alpha, \alpha \in \mathbb{N}. \\
\end{aligned}
\end{equation}
Here $x_{R}$ refers to the right end point of $\Omega_\C$. $b>1$ indicates how the grid spacing is increased and $a>0$ is a prefactor. 
Similarly, we can select such grid points in the bath on the left, {\it e.g.}, $x_\alpha = x_L -  a b^{-\alpha}, \alpha \in \mathbb{N}$.  Such grid distribution is known as the logarithmic grids in the solution of the Schr\"{o}dinger equation in polar or spherical coordinates.
Fig. \ref{fig: schematic2} demonstrates the distribution of such grid points, where the system consists of 80 Lithium atoms. It also shows a scenario of how the grid points can be selected. This approach assumes that the system is one-dimensional. But it can also be extended to systems that  are quasi one-dimensional, {\it e.g.}, a nanowire. 

By including these additional grid points, we introduce the computational domain
as the union of the selected grid points with the center region $\Omega_\C$, 
\begin{equation}
    \Omega_\i:=\Omega_\C \cup \Big\{x_\alpha, \quad \alpha \in \A\Big\}. 
\end{equation}
The symbol $\A$ indicates the labels of those grid points in the bath that have been selected according to \eqref{eq: log-grid}. The dimension of the new domain $\Omega_\i$ will be denoted by $m.$ Of our particular interest is  the case when $m \ll N.$
The new `bath' region corresponds to the complement of $\Omega_\i$, 
\begin{equation}
    \Omega_\ii:= \Omega\backslash\Omega_\i.
\end{equation}

\subsection{Galerkin projection for the Perturbed dynamics}


Motivated by its remarkable success of reduced-order modeling  techniques \cite{antoulas2010interpolatory,freund2000krylov,bai2002krylov} in handling large-dimensional dynamical systems, 
we apply the Petrov-Galerkin projection approach directly to the {\it full} system Eq.  \eqref{eq: drhoeq}. The method uses two subspaces $\V$ and $\W$ as the solution and test spaces, respectively. Although conventional Petrov-Galerkin projections arise from approximations of PDEs \cite{larsson2008partial}, the technique can also be applied to discrete dynamics, {\it e.g.}, classical particle systems \cite{li2014coarse,ma2019coarse}. For reduced-order problems, the approach is more robust than the traditional moment matching methods \cite{bai2002krylov}.
To facilitate implementations, the space $\V$ will be represented by a set of basis vectors, which are arranged into the columns of a matrix, denoted by $V$. Therefore, $\V=\text{Range}(V)$, and the same notations will be used for the space $\W$ as well: $\W=\text{Range}(W)$.    The dimension of the two subspaces is assumed to be $m$.

\medskip 

Overall,  the Petrov-Galerkin projection for  Eq  \eqref{eq: drhoeq}  consists of the following three steps:
\begin{enumerate}
    \item [{\bf (1)}] Seek the solution in the subspace $\V$. In our case, the approximate density matrix is spanned by the matrix $V$, as follows,
\begin{equation}\label{eq: delrho}
\drho (t) \approx \dwrho(t) = V D(t)V^{*}.
\end{equation}
In the subspace approximation, the $m\times m$ matrix $D$ will be referred to as the {\it nodal values}. 

\item [{\bf (2)}]   Project the dynamics \eqref{eq: drhoeq} to the orthogonal complement of a subspace $\W$. We choose the subspace spanned by the column vectors of a matrix $W$, which yields, 
\begin{equation}\label{eq: proj}
	i\frac{d}{dt} W^{*}\dwrho(t) W= W^{*} \big( [H(t),\dwrho(t)] + \Theta(t) \big) W.
\end{equation}
The first two steps offer a low-rank approximation of the density-matrix. Such projections have been widely used in solving matrix equations \cite{simoncini2009convergence}.

\item [{\bf (3)}]  Use quadrature formulas to approximate functions  outside the domain $\Omega_\i$ using the nodal values in $D(t)$.

\end{enumerate}

The matrix products in the first two steps of the Petrov-Galerkin projection give rise to two coefficient matrices,
\begin{equation}\label{eq: MK}
    {M} =V^*W, \quad {K} = V^{*}H W. 
\end{equation}
In the Galerkin projection, these matrices are often referred to as the mass and stiffness matrices \cite{larsson2008partial}. Direct computations from the first step \eqref{eq: delrho} and second step \eqref{eq: proj}  yield,
\begin{equation}\label{eq: Dequation}
	i\frac{d}{dt} D(t) = [{H}_\text{eff},D] + {\Theta}_\text{eff}(t),
\end{equation}
where ${H}_\text{eff}$ is the reduced {Hamiltonian},
\begin{equation}\label{eq: tildeH}
	{H}_\text{eff} = {K} {M}^{-1}. 
\end{equation}
The bracket \eqref{eq: Dequation} has been generalized to non-Hermitian matrices,
\begin{equation}
    [A, B]:= A^*B - B^*A.
\end{equation}

Furthermore, the non-homogeneous term  ${\Theta}_\text{eff}(t)$ is a reduction of the non-homogeneous term $\Theta(t)$ in \eqref{eq: drhoeq}: 
\begin{equation}\label{eq: theta'}
 {\Theta}_\text{eff}(t) = M^{-*}W^{*} \Theta(t) WM^{-1} = M^{-*}W^{*}[\dH(t), \rho_{0}]W M^{-1}.
\end{equation}
This matrix product will be treated in step {\bf (3)}, {\it i.e.},  approximated by a quadrature formula. 

\subsection{The choice of the subspaces}

We will choose the subspace $\mathbb{V}$ by specifying a set of orthonormal basis following our previous work \cite{chu2020reduced}. To be more specific, using the partition,
 $    \Omega = \Omega_\i \cup \Omega_\ii,$
we express the basis as columns of a matrix $V$,
\begin{equation}\label{eq: V}
   {V} = \left[\begin{array}{c}
         I_{m\times m}  \\
         0 
    \end{array}\right]\!\!
    \begin{array}{c}
         \Omega_\i  \\
         \Omega_\ii 
    \end{array}. 
\end{equation}
Here we follow the notations in domain decomposition methods \cite{quarteroni1999domain}, and order the grid points so that the points in $\Omega_\i$ appear first.  Thus, in \eqref{eq: V}, we pick the standard basis in the subdomain $\Omega_\i.$ Then they are extended to  $\Omega_\ii$ by filling in zeros. This choice is motivated by
the reduced-order modeling \cite{bai2002krylov}. See  \cite{chu2020reduced} for details about how the density-matrix equation \eqref{eq: drhoeq} can be 
put into a standard form of   reduced-order problems. In general, the choice of $V$ corresponds to the identification of quantities of interest by defining the output vector \cite{bai2002krylov}. As we will demonstrate later, it will single out the components of the density matrix in the subdomain $\Omega_\i$, leading to a description using the reduced density matrix (RDM). In turn, one may consider how the rest of the dynamics is controlled by the RDM.  The vectors in $W$  represent such coupling. Specifically, we choose, 
\begin{equation}\label{eq: W}
   { W} = \big(\varepsilon I - H[n_0]\big)^{-1} {V}.  
\end{equation}
Here we require $\text{Im}(\varepsilon)<0$ for stability \cite{chu2020reduced}.  

With the choices \eqref{eq: V} and \eqref{eq: W}, one finds that,
\begin{equation}
     M= V^* W= V^*(\varepsilon I - H[n_0])^{-1} V = G_{\i,\i}(\ve). 
\end{equation}
Here $G_{\i,\i}$ refers to the first diagonal block of the Green's function  \cite{williams1982green,economou2006green},
$G(\ve)$. From the block inversion formula, we obtain,
\begin{equation}\label{eq: matM}
    M^{-1}= \varepsilon I  - H_{\i,\i} - H_{\i,\ii} \big[\varepsilon I - H_{\ii,\ii} \big]^{-1} H_{\ii,\i}.  
\end{equation}

Meanwhile, for the stiffness matrix, we have 
\begin{equation}
K= V^* H W= - \ve V^*V + \ve M.
\end{equation}
Combining with \eqref{eq: tildeH} and \eqref{eq: matM}, and noticing the orthogonality 
$V^*V=I$, we get,
\begin{equation}
    	{H}_\text{eff} = H_{\i,\i} + \Sigma_{\i,\i}, 
\end{equation}
where $\Sigma_{\i,\i}$ is the self-energy \cite{brandbyge2002density}, given by
\begin{equation}
\Sigma_{\i,\i}  =  	H_{\i,\ii} \big[\varepsilon I - H_{\ii,\ii} \big]^{-1} H_{\ii,\i}.
\end{equation}
Therefore the choices of the subspaces according to \eqref{eq: V} and \eqref{eq: W} have a remarkable connection with the well known Green's function approach \cite{williams1982green,KellyCar92}. In particular, the self-energy $\Sigma_{\i,\i}$ describes the influence from the domain $\Omega_\ii$. It is also worthwhile to point out that since $\Omega_\i$ is multi-connected, as depicted in Fig. \ref{fig: schematic2}, the self-energy $\Sigma_{\i,\i}$ is an operator that couples a domain with many fragments. 

\subsection{The computation of the Coulomb potential}\label{sec: alg1}

The selection of the domain $\Omega_\i$ is with the purpose to take into account the density fluctuation $\dn$ in the bath. With the electron density in the domain $\Omega_\i$, we can reconstruct the correction to the Hartree part of the Hamiltonian \eqref{eq: vH} using a quadrature formula,
\begin{equation}\label{eq: Q}
    V_H[\dn](x) = \int_{\mathbb{R}} v(x-x') \dn(x') dx' 
\approx  \sum_{x'\in \mathscr{Q}_1} w(x,x') \dn(x'). 
\end{equation}
Here $\mathscr{Q}_1$ denotes the set of quadrature points and $w(x,x')$ is the quadrature weights. In the numerical tests, we will use the 3-point Gaussian quadrature for each interval in $\Omega_\i$. 

Since $x_\alpha$ from the bath is included in the computational domain $\Omega_\i$, the density-matrix at $x_\alpha$ is explicitly computed, and as its diagonal elements, $\dn(x_\alpha)$ is accessible. In the computation, we use a spline interpolation based on the electron density at these points, which is then evaluated at those quadrature points $x\in \mathscr{Q}_1$, so that the quadrature formula \eqref{eq: Q} can be applied to update the Hartree potential. 

\subsection{The evaluation of $\Theta_\text{eff}$}\label{sec: alg2}

We now discuss how the non-homogeneous term \eqref{eq: theta'} is evaluated. Due to the dependence of $\dH$ on the electron density \eqref{eq: dH}, this term has to be computed at each time step.  
The main issue here is that the matrix product with $W$ involves a summation over the {\it entire} domain, which requires $\mathcal{O}(m^2 N^2)$ operations.  

As with many other applications of Galerkin projection \cite{strang1973analysis,larsson2008partial}, this term can be treated by quadrature approximations.  For convenience, let us first work with a summation form. Namely, we consider,
\[ I[f] = \sum_{x_k \in \Omega}  f(x_k).\] 
In the present setting, $x_k$'s are uniformly distributed grid points in the entire domain.
We assume that $f(x)$ has certain smoothness ({\it e.g.} twice continuously differentiable). In the quadrature approximation, the sum is approximated by,
\[ I[f] \approx \sum_{x_\alpha \in \mathscr{Q}_2 } f(x_\alpha) A_\alpha.\]
Here $\mathscr{Q}_2$ denotes a subset of the points, which will be regarded as quadrature points and $A_\alpha$ will be the corresponding weights.
The quadrature weights in our computation are determined in advance by a spline interpolation. A simple idea is the following: For each $x_\alpha$, one can set $f(x_\alpha)=1$ but zero for all other quadrature points. Using a cubic spline interpolation, $f(x) \approx s(x),$  and from the quadrature formula above, one can obtain the quadrature weight: $A_\alpha = I[s],$ {\it i.e.}, directly summing up the spline interpolant $s(x)$.

Returning to \eqref{eq: theta'}, with the quadrature formula, we can reduce the matrix product,
\begin{equation}\label{eq: th-quad}
\begin{aligned}
\left( W^{*}\big[\dH, \rho_{0}\big]W\right)_{ij} &=
\sum_k \sum_\ell \overline{W_{k,i}} ( \dH_k -\dH_\ell ) \rho^{0}_{k,\ell}     W_{\ell, j}, \\
&\approx  \sum_{x_\alpha \in \mathscr{Q}}  \sum_{x_\beta \in \mathscr{Q}} A_\alpha A_\beta  \overline{W_{\alpha,i } } ( \dH_\alpha  -\dH_\beta  ) \rho^{0}_{\alpha,\beta}     W_{\beta, j}.
\end{aligned}
\end{equation}
The overline denotes the complex conjugate. If the number of quadrature points is $k$, this procedure reduces the computation from $\mathcal{O}(m^2 N^2)$ 
to $\mathcal{O}(m^2 k^2)$. 

An interesting case is when $\mathscr{Q}_2=\Omega_\i$. Combining \eqref{eq: theta'} and \eqref{eq: th-quad}, we have,
 \begin{equation}
     M^{-*}W^{*}\big[\delta V, \rho^{(0)}\big]WM^{-1}  \approx  M^{-*}W^{*} V^T S V W M^{-1} = S,
 \end{equation} 
where $S$ is a skew-Hermition matrix with elements given by, 
\begin{equation}\label{eq: Smat}
S_{\alpha,\beta}= A_\alpha A_\beta  ( \delta V_\alpha  -\delta V_\beta  ) \rho^{0}_{\alpha,\beta}. \end{equation}
In the last step, we have used the equation \eqref{eq: MK}.

\begin{algorithm}[H]
		\caption{Implementation of Eq. \eqref{eq: Dequation} }
		\begin{algorithmic}
	\State $t=t_n, D= D(x, x', t_n)$
	\For {$x_\alpha \in \Omega_\i$}
	\State $\dn(x_\alpha)= D(x_\alpha,x_\alpha)$
    		\State $\dV_{XC}= V_{XC}[n_0 + \dn] - V_{XC}[n_0] $
 \EndFor
\State Interpolate $\dn$ and $\dV_{XC}$ using cubic spline functions, $\wt{\dn}$ and $\wt{dV}_{XC}$, respectively.  
\For {$x_\alpha \in \Omega_\i$}
\State  Compute the values of $\dn(x)$ at the quadrature points  from  $\wt{\dn}$.
	\State Compute the Coulomb potential $V_H[\dn](x_\alpha)$ from \eqref{eq: Q} using Gaussian quadrature.
	\For {$x_\beta \in \Omega_\i$}
	\State Compute the matrix elements $\Theta_\text{eff}$ from \eqref{eq: th-quad}.
	\EndFor
	\EndFor
	\State Use $H_\text{eff}$ and $\Theta_\text{eff}$ to update the density matrix $D$ to the next time step $t_{n+1}$.
	\end{algorithmic}
	\label{alg}
\end{algorithm}

\subsection{Methods for the Time Integration} \label{sec: alg3}
In preceding sections, we explained how the terms in the RLvN equation \eqref{eq: Dequation} are computed at each time step, and they   are summarized in  Algorithm \ref{alg}. Here we discuss how the dynamics is integrated. 

We first discuss the approximation of the full model \eqref{eq: lvn}. We assume that a finite-difference approximation has been applied so that 
the Hamiltonian operator, as well as the charge density, are represented at grid points.  We adopt an operator-splitting scheme by Watanabe and  Tsukada \cite{watanabe2002efficient} for TDDFT models. More specifically, we split the Hamiltonian as follows,
\begin{equation}
    H = H_1 + H_2, \quad H_1:=-\frac12{\nabla^2}, \quad H_2:= H-H_1,
\end{equation}
where the potential term may include Coulomb, exchange-correlation, and external biased potential. 

Despite the non-linearity in \eqref{eq: lvn}, one can still express the solution with an evolution operator,
\begin{equation}
    \rho(t+\dt)= \E(t+\dt,t) \rho(t).
\end{equation}
The evolution operator can be effectively approximated by an operator-splitting scheme,
\begin{equation}\label{eq: exp}
    \E(t+\dt,t) \approx \E_1(t+\dt,t+\frac{\dt}2)\E_0(\dt)\E_1(t+\frac{\dt}2,t).
\end{equation}

Specifically, the operator $\E_0$ corresponds to the dynamics determined by the kinetic energy, 
\begin{equation}
    i \partial_t \rho = [H_1, \rho],
\end{equation}
for which the evolution operator corresponds to a unitary dynamics:
\begin{equation}
  \E_0(\dt) \rho = U_0(\dt) \rho U_0(\dt)^\dagger, \quad U_0(\dt):= \exp (-i \dt H_1 ).
\end{equation}
The matrix exponential in $U_0$ can be computed from direct spectral decomposition, {\it e.g.}, by using fast Fourier transform. 

On the other hand, the operator $\E_1$ dictates the dynamics,
\begin{equation}
    i \partial_t \rho = [H_2, \rho].
\end{equation}
The diagonal of the right hand side is zero, which implies that the electron density remains constant. Therefore, this is effectively a {\it linear} equation. As a result, we can write,
\begin{equation}
    \E_1(t+\dt,t) \rho(t) = U_1(\dt) \rho(t)  U_1(\dt)^\dagger, \quad U_1(\dt):= e^{-i\dt H_2}.
\end{equation}
In addition, since $H_2$ is only diagonal, the evaluation is straightforward.

\medskip

We now turn to the RLvN \eqref{eq: Dequation}. Using the variation-of-constant formula, we first obtain an integral form of the solution (with the spatial variable $x$ and $x'$ suppressed),
\begin{equation}\label{eq: vacf}
    D(t+\dt) = U(\dt) D(t) U(\dt)^\dagger - i \int_0^{\dt} U(\dt - \tau) \Theta_\text{eff} (t+\tau ) U(\dt - \tau)^\dagger d\tau.
\end{equation}
Here $U(\dt) = \exp \big( -i \dt H_\text{eff} \big),$ which only needs to be computed once.  Since the term $\Theta_\text{eff}$ in the equation \eqref{eq: theta'} also depends on the diagonals of $D(t)$, this remains as an integral equation. To obtain an approximation, one can approximate the integral using the values at $t$ (\(\tau =0\) ). This yields,
\begin{equation}
    D(t+\dt) = U(\dt) \Big( D(t) - i \dt \Theta_\text{eff} (t) \Big)   U(\dt)^\dagger,
\end{equation}
which is an analogue of the exponential-Euler method \cite{minchev2005review}.

This method can be extended to form a class of integrators that resemble the exponential-Runge-Kutta methods \cite{hochbruck2005explicit}. As an example,  one can apply the above formula for half of a step, which is then combined with  the mid-point rule for the integral \eqref{eq: vacf}:
\begin{equation}\label{eq: exp-rk2}
    \begin{aligned}
       D(t+\frac{\dt}2) &= U(\frac{\dt}2) \Big( D(t) - i \frac{\dt}2 \Theta_\text{eff} (t) \Big)   U(\frac{\dt}2)^\dagger, \\
        D(t+\dt) &= U(\dt) \Big( D(t) - i \dt \Theta_\text{eff} (t+\frac{\dt}2) \Big)   U(\dt)^\dagger.
    \end{aligned}
\end{equation}
When the numerical stability is largely determined by the spectrum of $H_\text{eff}$, such exponential integrators often allow much larger step sizes.

\section{Numerical tests}\label{sec: II}

As examples, we consider a 1d Lithium (Li) system, consisting of three Li chains with lattice spacing 2.8\AA. Two atom chains, each with 78 atoms, are used to model the bath. They are separated from  the middle chain with 4 Li atoms by adding an extra spacing of 2.8\AA. The system is larger than the one depicted in Fig. \ref{fig: schematic2}. We choose a relatively large system so that the full system can still be simulated within a realistic time frame to generate a reference solution. 
 Similar systems have been used by \cite{ke2007role,thakur2015electronic} to study  transport properties.

The units follow those in \cite{baker2015one}, using Bohr radius and eV as the length and energy units, which subsequently determine the time units. 
We use a real-space method to approximate the solution of  the full LvN model \eqref{eq: lvn} as well as the reduced model that follows. Specifically, we use a finite-difference spatial discretization with $N=4000$ grid points and grid spacing $\Delta x= 0.1134$. We use the 3-point finite difference formula for the kinetic energy term. The electron current can be extracted from the off-diagonal entries of the density matrix,
\[J(x_j)= - \text{Im}(\rho(x_j,x_{j+1}))/\Delta x.\]  
The initial ground state is computed self-consistently using a mixing method \cite{anderson1965iterative,pulay1980convergence} with $4$ previous iterations at each step. The step size for the time integration is $\dt=0.00125$ and the equations \eqref{eq: lvn} and  \eqref{eq: Dequation} are integrated up to $t=16.$  Due to the finite size of the bath, boundary effect will occur. But we have verified that such finite-size effect will not affect the dynamics in the center region within this time period. 

\begin{figure}[H]
    \centering
    \includegraphics[width=0.5\textwidth]{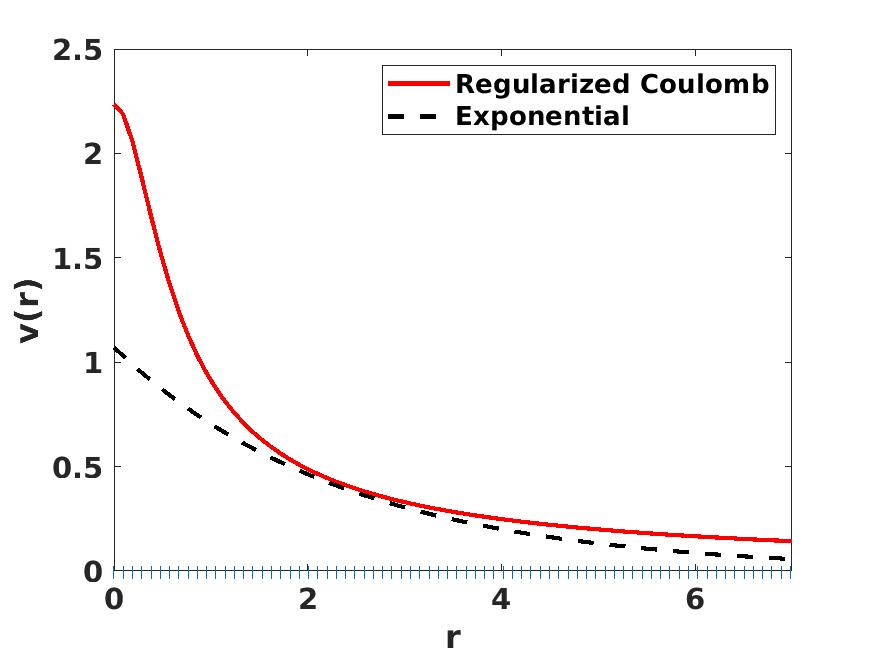}
    \caption{ The long-range interactions: Regularized Coulomb (solid line); Exponential potential from \cite{baker2015one}. To show scales, the grid points are shown at the bottom.
}
    \label{fig: cpot}
\end{figure}

We follow  the model by Baker et al. \cite{baker2015one} for the Coulomb potential and  the exchange and correlation potentials. 
Fig. \ref{fig: cpot} shows the exponential  form from \citep{baker2015one} for the Coulomb potential, plotted together with the grid points in the real-space approximation to show the scale. Also shown in the figure is the regularized Coulomb potential (Yukawa):
\begin{equation}\label{eq: yukawa}
    v(r)=1/\sqrt{r^2+0.2}.
\end{equation}
 In terms of the atomic spacing, the exponential model has an effective range up to second nearest neighbors. But the regularized Coulomb potential extends much farther. Both models will be implemented in the computation.

For the reduced model \eqref{eq: Dequation}, we first pick the subdomains as follows. The entire domain is positioned in the interval
$[0,L]$ with $L=453.6.$ We keep the center region in the interval $[x_L,x_R]=[205.8, 247.8].$ Starting from this center region, we gradually increase the grid size. For example, in the bath on the right, we define the logarithmic grid  
\(x_\alpha= x_R + \lfloor 2\times 1.2^{\alpha}\rfloor \Delta x\) with $\lfloor\cdot\rfloor$ denoting the nearest integer. Out of the 4000 grid points,  $m=462$ points have been selected with this procedure. 
3-point Gaussian quadrature formulas are used for the calculation of the Coulomb potential  in \eqref{eq: Q}. For the non-homogeneous term \eqref{eq: theta'},  since we are approximating a summation, we need to pick quadrature points that coincide with the grid points. In the center region, we already keep all the grid points. Meanwhile,  in the bath, for each interval with  $x_{\alpha+1} - x_\alpha >  9\Delta x $, we introduce two additional quadrature points in that interval.

\begin{figure}[H]
    \centering
    \includegraphics[width=0.65\textwidth]{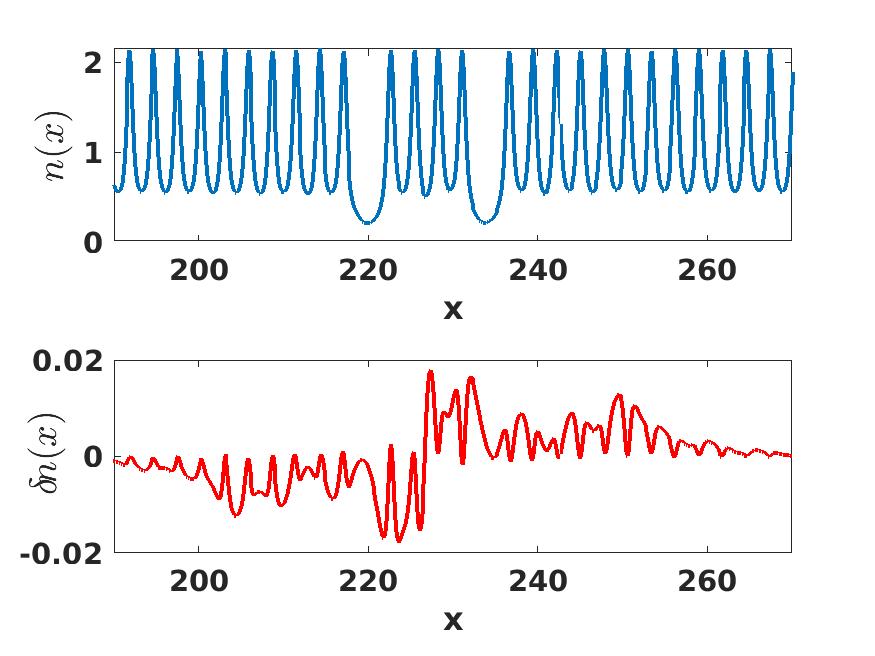}
    \caption{Top: The charge density. Bottom: The fluctuation of the density $\dn$.
}
    \label{fig: density}
\end{figure}

Using the operator-splitting method \eqref{eq: exp}, we first run the full model \eqref{eq: lvn} up to $t=16.$ 
 We exert a potential bias $U_L=0.1$ for $x<L/2$ and $U_R=0$ in the other half of the region.  We first show the charge density at $t=8$, along with the charge fluctuations $\dn(x,t)$ around the center region in  Fig. \ref{fig: density}. We observe that the charge density $n(r,t)$ exhibits oscillations around the nuclei in the bath which extends to the entire domain. The two 'dips' in the density profile occur at the junctions.  In contrast, the charge perturbation $\dn(x,t)$ exhibits decay toward the far side of the bath. The reduced model  works with this charge perturbation, which is represented at sparse grid points with increasing spacing. 

\begin{figure}[H]
    \centering
    \includegraphics[width=0.6\textwidth]{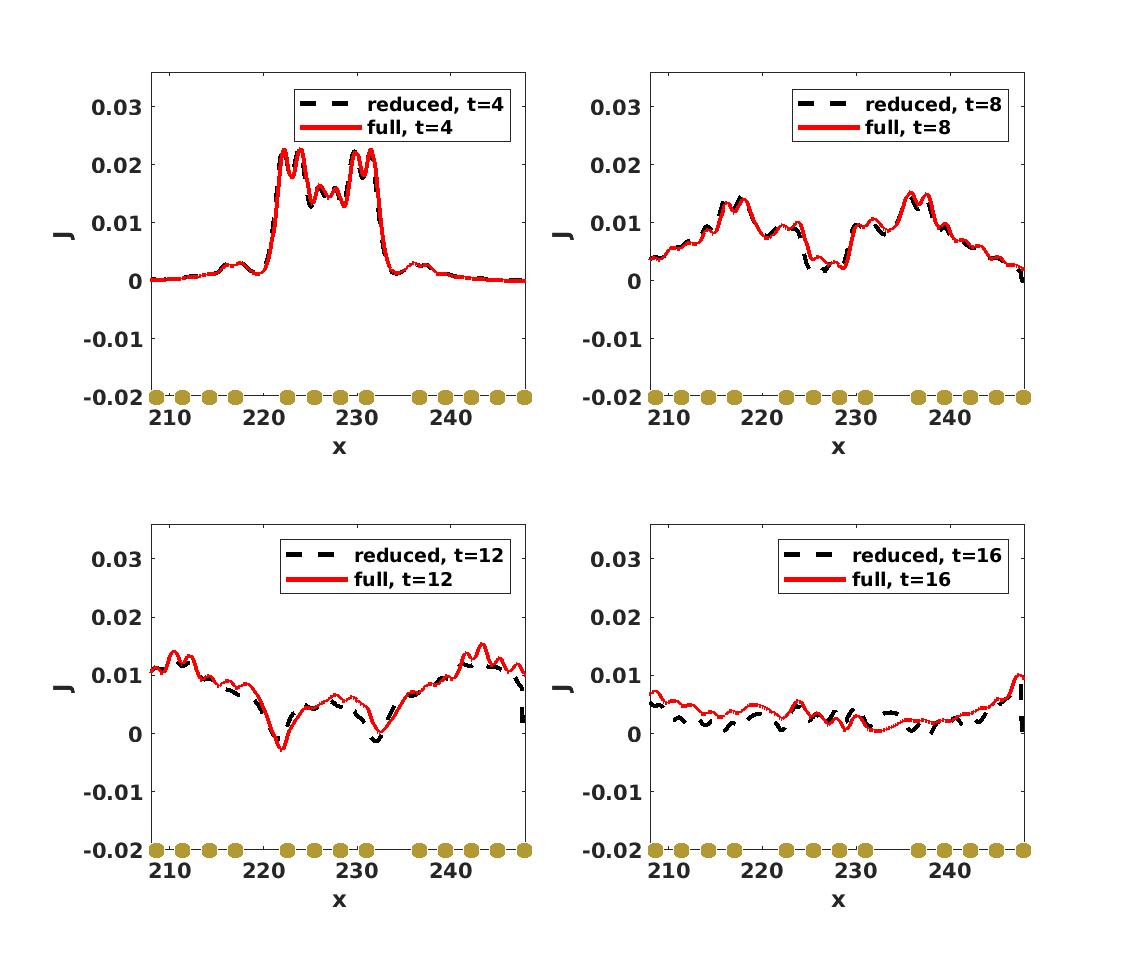}
    \caption{Comparison of the computed current around the junction for various points in time. As a reference, the position of the atoms in the region is plotted at the bottom of the panels. 
}
    \label{fig: currX}
\end{figure}

\begin{figure}[H]
    \centering
    \includegraphics[width=0.55\textwidth]{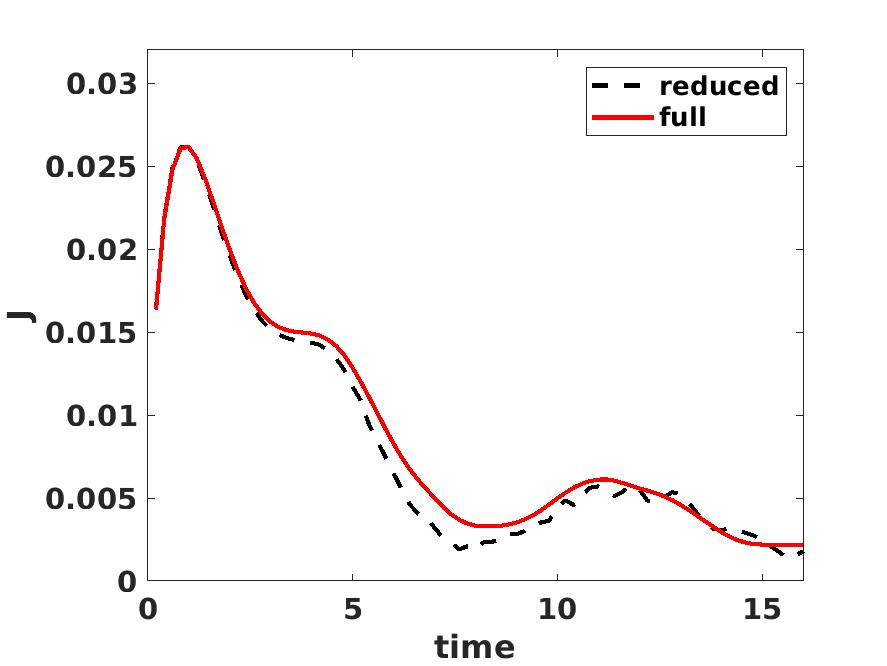}
    \caption{Comparison of the computed current at the center $x=L/2$ for the time window $[0,16]$.
}
    \label{fig: currT}
\end{figure}

Next we implement the algorithm \eqref{eq: exp-rk2} and run the reduced model \eqref{eq: Dequation} using the exponential Coulomb potential. For the subspace $\W$ \eqref{eq: W}, we pick the shift to be $\varepsilon= \mu - 5 i$, with $\mu$ being the Fermi level. Fig. \ref{fig: currX} depicts the electron current around the junction for various time instances. One can observe that the potential bias creates a current peak initially, which then develops into an oscillatory pattern. Subsequently, the current profile  starts to widen and spread toward the bath region.  The time history of the current at $x=L/2$ is shown in Fig. \ref{fig: currT}. In both cases, the results agree well with the full simulation results. 

\begin{figure}[H]
    \centering
    \includegraphics[width=0.6\textwidth]{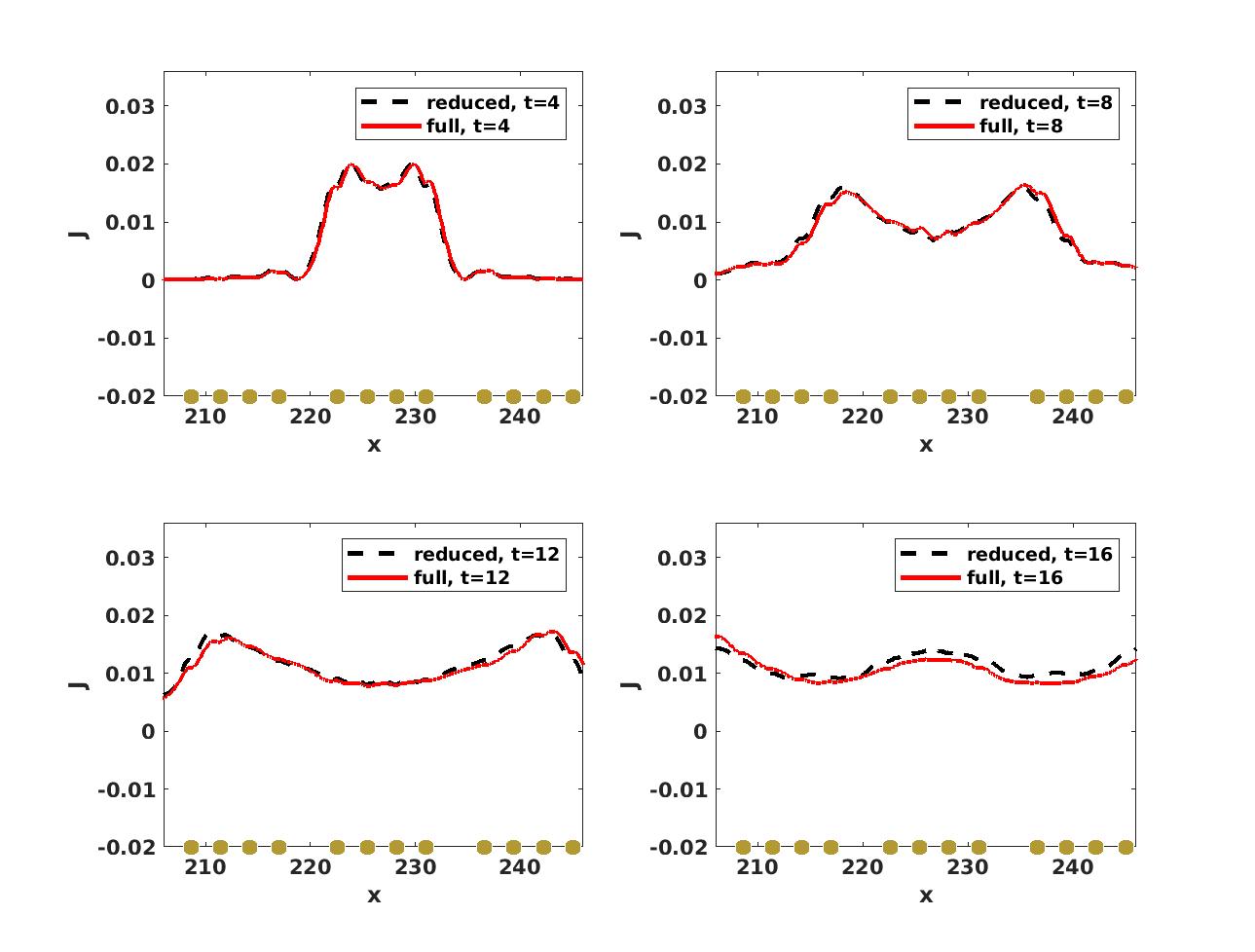}
    \caption{Comparison of the computed current around the junction for various points in time with the regularized Coulomb interaction \eqref{eq: yukawa}. As a reference, the position of the atoms in the region is plotted at the bottome. 
}
    \label{fig: currXc}
\end{figure}

\begin{figure}[H]
    \centering
    \includegraphics[width=0.55\textwidth]{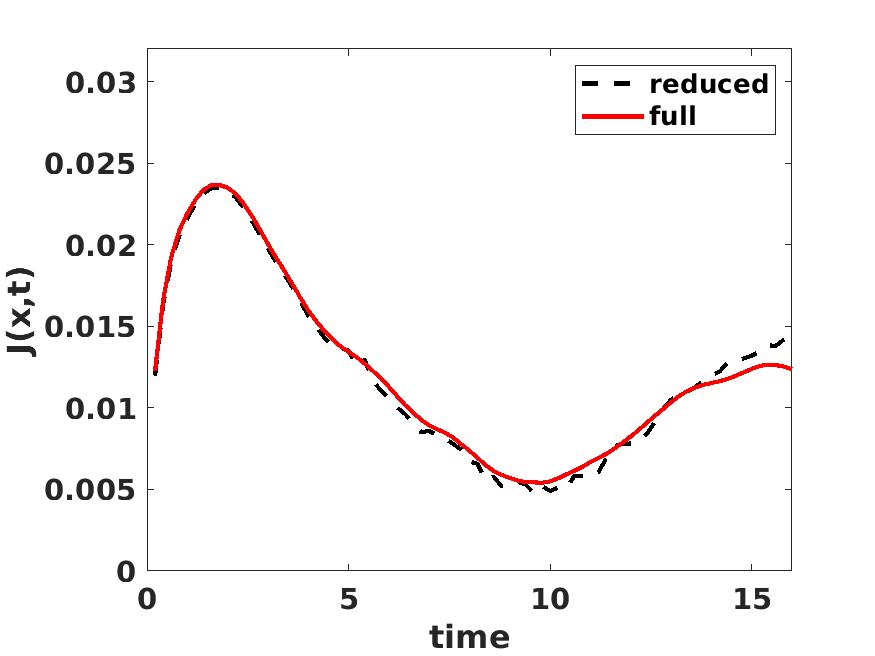}
    \caption{Comparison of the computed current at the center $x=L/2$  with the regularized Coulomb interaction \eqref{eq: yukawa} over the time period $t\in [0, 16]$.
}
    \label{fig: currTc}
\end{figure}

The numerical tests are repeated using the regularized Coulomb interaction \eqref{eq: yukawa}, in which case, the Coulomb potential at the junction has a stronger dependence on the electron density in the bath. Compared to the previous case, the current profile is more smooth, as shown in Fig. \ref{fig: currXc}. In this case,  the reduced models still accurately predicted the current in the center region. The comparison  of the current at $x=L/2$ is shown in 
Fig. \ref{fig: currTc}. 

It is also worthwhile to point out that the reduced-order approach \cite{antoulas2010interpolatory,bai2002krylov} is effective in predicting pre-selected quantities of interest (the output),  which in our case, is the density-matrix in the  region $\Omega_\i$ and it embodies the electron current in the center region. However, it does not involve the current in the bath regions. Therefore, the reduced model  \eqref{eq: Dequation} will not be able to capture the current outside the center region. 


Finally, we monitored the CPU time for solving the full LvN equation \eqref{eq: lvn} and the RLvN \eqref{eq: Dequation}. We run the models for systems of various sizes. The results are summarized in Table \ref{tab:cpu}. As the dimension $N$ increases, the CPU time corresponding to the simulation of the full LvN equation \eqref{eq: lvn} increases considerably. The CPU time for the RLvN \eqref{eq: Dequation} follows a similar scaling but it remain remarkably low. 
\begin{table}[thp]
\caption{ CPU time (in seconds) comparison for 100 steps of time integration \label{tab:cpu}
}
 \begin{tabularx}{\textwidth}  {XXXX}
 \hline\hline
 & $ N=2400$ & $N=3200$ & $N=4800$ \\ 
 \hline
  Full  LvN & 3103 & 8694 & 25409\\
  Reduced LvN & 63 &114  & 312 \\
  \hline\hline
 \end{tabularx}
 \end{table}

\section{Summary}
In this paper, we have constructed a class of reduced-order models for simulating electron transport problems. Based on a decomposition of the physical domain, and by partitioning  the density matrix accordingly into different blocks, we regard such a dimension reduction problem as a reduce-order problem  \cite{chu2020reduced}, where a subspace projection method can accurately predict  certain quantities of interest  \cite{freund2000krylov,bai2002krylov}.  This approach naturally involves the self-energy in the effective Hamiltonian. In this work, our focus is placed on  the treatment of long-range interactions. Specifically, we choose subspaces so that the electron density in the bath is captured on some sparse grids, which in turn determines the Hartree potential. This enables the accurate computation of the Coulomb and exchange-correlation potentials at the molecular junction.  It is possible to enlarge the subspaces to enable more accurate approximations, as demonstrated in our prior work \cite{chu2020reduced} for a tight-binding model. But based on the current results, we observed that the present approach using the subspaces \eqref{eq: V} and \eqref{eq: W} is quite easy to implement, and it has already shown good agreement with the full model. 
\begin{acknowledgments}
Li's research is supported by NSF under grant DMS-1819011 and DMS-1953120. 
\end{acknowledgments}

\bibliography{etransp}
\end{document}